\documentclass{emulateapj}
\usepackage{amsfonts}
\usepackage{amssymb}
\usepackage{natbib}
\shorttitle{Dust emission from the parsec-scale in NGC 4151}
\slugcomment{accepted by ApJL}
\begin{document}
\title[Dust emission from the parsec-scale in NGC 4151]{Dust emission from a parsec-scale structure in the Seyfert 1 nucleus of NGC 4151\footnote{Based on observations collected at the European Southern Observatory, Chile, programme number 081.B-0092(A).}}
\author{L. Burtscher$^{1,2}$, W. Jaffe$^3$, D. Raban$^3$, K. Meisenheimer$^1$, K. R. W. Tristram$^4$, H. R\"ottgering$^3$}
\affiliation{$^1$ Max-Planck-Institut f\"ur Astronomie, K\"onigstuhl 17, 69217 Heidelberg, Germany\\
\email{burtscher@mpia.de}
$^2$ Fellow of the International Max Planck Research School (IMPRS) for Astronomy and Cosmic Physics, Heidelberg, Germany\\
$^3$ Sterrewacht Leiden, Leiden University, Niels-Bohr-Weg 2, 2300 CA Leiden, The Netherlands\\
$^4$ Max-Planck-Institut f\"ur Radioastronomie, Auf dem H\"ugel 69, 53121 Bonn, Germany}
\keywords{techniques: high angular resolution --- galaxies: active --- galaxies: Seyfert}
%
%
\newcommand{\ngc}{\ensuremath{\mathrm{NGC}\,4151}}
\begin{abstract}
We report mid-IR interferometric measurements with $\sim$ 10 mas resolution, which resolve the warm (T = $285^{+25}_{-50}$ K) thermal emission at the center of NGC 4151. Using pairs of VLT 8.2 m telescopes with MIDI and by comparing the data to a Gaussian model, we determined the diameter of the dust emission region, albeit only along one position angle, to be ($2.0 \pm 0.4$) pc (FWHM). This is the first size and temperature estimate for the nuclear warm dust distribution in a Seyfert 1 galaxy. The parameters found are comparable to those in Seyfert 2 galaxies, thus providing direct support for the unified model. Using simple analytic temperature distributions, we find that the mid-infrared emission is probably not the smooth continuation of the hot nuclear source that is marginally resolved with K band interferometry. We also detected weak excess emission around 10.5 $\mu$m in our shorter baseline observation, possibly indicating that silicate emission is extended to the parsec scale.
\end{abstract}
%
%
%
\maketitle
\section{Introduction}
In the standard model of Active Galactic Nuclei (AGNs) \citep{antonucci1993,urry1995}, the central engine (Black Hole + accretion disk + Broad Line Region) is embedded in a dusty torus. Galaxies where we see the center only through this torus are called ``type 2'' whereas ones where we receive direct radiation from the Broad Line Region are referred to as ``type 1''. For a long time, there was only indirect evidence for the existence of the enshrouding dust, until VLTI observations with MIDI actually resolved the parsec-scale AGN heated dust structures in two Seyfert 2 galaxies (Sy 2). \citep{jaffe2004,tristram2007b,raban2009}

Further observations proved the existence of parsec sized dust structures in several Seyfert 1 galaxies (Sy 1) \citep{tristram2009}, but so far, no Sy 1s have been observed in sufficient detail to test the central premise of the unified models: type 1 and type 2 dust distributions are identical and the observed differences are only due to differences in orientation with respect to the line of sight. These predictions can now be tested with the MIDI observations of NGC~4151 reported here.

At a distance $D = (14 \pm 1)$ Mpc (i.e. 1~mas $\sim$ 0.068~pc)\footnote{from the NASA Extragalactic Database: http://nedwww.ipac.caltech.edu; distance from redshift with H$_0$=73 km/s/Mpc; other estimates range to 20 Mpc}, NGC~4151 is the closest and brightest type 1 galaxy (classification: Seyfert 1.5). It is also one of the most variable Seyfert galaxies: The UV continuum flux varies on scales of days and weeks \citep{ulrich2000} and the reverberation time to the hot dust on the sub-parsec scale varies on yearly timescales \citep{koshida2009}.

Sy 1 galaxies have been observed previously with MIDI \citep[NGC~3783,][]{beckert2008}, \citep[NGC~7469,][]{tristram2009}, but NGC~4151 is the first multi-baseline case where the size of the nuclear dust distribution is clearly indicated.

\citet{swain2003} reported near-IR, 2.2$\mu$m, observations with the Keck interferometer. They find that the majority of the K band emission comes from a largely unresolved source of $\le 0.1$ pc in diameter. Based on this small size, they argued that the K band emission arises in the central hot accretion disk. They note, however, that their result is also consistent with very hot dust at the sublimation radius. This view is supported by the K band reverberation measurements of \citet{minezaki2004} who find a lag time corresponding to $\sim 0.04$ pc.

\citet{riffel2009} modelled the near-infrared spectrum of this source and found it to be composed of a powerlaw accretion-disk spectrum and a component likely arising from hot dust (T = 1285 K). The hot dust component dominates at $\lambda \gtrsim 1.3 \mu$m, consistent with the interpretation of hot dust emitting at $2\mu$m. They measured a K band flux of $\sim 65$ mJy.

In the mid-infrared, \citet{radomski2003} presented images of NGC~4151 at 10.8$\mu$m with an aperture of 4.5\arcsec\, and a spatial resolution of $\sim$550 mas (35 pc). They find that the majority (73$\%$) of the N band flux comes from an unresolved point source with a size $\le35$pc, and the rest is extended emission from the narrow line region.
 
In this letter we report new mid-infrared interferometric observations of NGC~4151 which clearly resolve a thermal structure.
\section{Instrument, observations and data reduction}
\label{sec:obs}

Observations were performed in the \emph{N} band ($\lambda \approx 8 \ldots 13 \mu\mathrm{m}$) with the MID-infrared Interferometric instrument \citep[MIDI,][]{leinert2003b} at the ESO Very Large Telescope Interferometer (VLTI) on Cerro Paranal, Chile, using pairs of 8.2\,m Unit Telescopes (UTs). The observables with MIDI are the single-dish spectra and the correlated flux spectrum that is obtained from the interference pattern generated by the two beams. The spectra were dispersed with a NaCl prism with $R\equiv \lambda/\delta\lambda \sim 30$.

Observations were taken in the nights of April 21 and 23, 2008 with projected baseline length $BL$ of 61\,m and 89\,m at position angles of 103$^{\circ}$ and $81^{\circ}$ respectively. These provide effective spatial resolutions\footnote{Applying Rayleigh's criterion to an interferometer leads to a resolution of $\lambda / 2 BL$. However, given sufficient S/N, one can already distinguish \emph{models} at lower resolutions, in our case at $\sim \lambda / 3 BL$.} ($\lambda / 3 BL$) at 10.3$\mu$m\, of 11 and 7 milliarcseconds (mas), respectively. The calibrators HD~133582 and HD~94336 were selected to be very close in airmass with $\Delta (\sec z) \lesssim 0.15$. This is especially important for NGC 4151 (DEC $\sim +40^{\circ}$) which, at Paranal, never rises higher than $\sim 25^{\circ}$ ($\sec z \gtrsim 2.3$) above the horizon. The northern declination of the source also essentially limits the projected baselines and fringe patterns to East-West orientation at Paranal. The N band spectrum of HD~133582 (K2III) was taken to follow a Rayleigh-Jeans law, while that of HD~94336 (MIII) was taken from \citet{cohen1999}.

Data reduction was performed with the interferometric data reduction software \emph{MIDI Interactive Analysis and Expert Work Station} \citep[MIA+EWS,][]{jaffe2004b}, specifically adapted for the analysis of low S/N sources.
\section{Results and modelling}
\label{sec:res}

\subsection{Single-dish spectrum}
\begin{figure}
	\begin{centering}
		\includegraphics[width=9.5cm,trim=3cm 2cm 0cm 3cm]{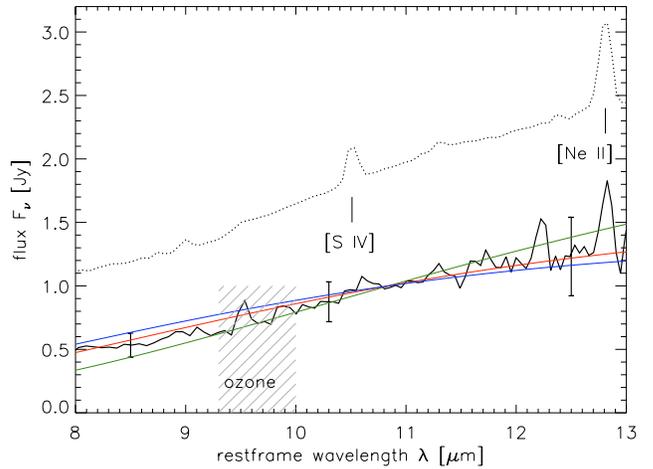}
		\caption{Single-dish spectra for NGC~4151. Spitzer spectrum \citep[3.5\arcsec, dotted,][]{weedman2005}, MIDI spectrum (0.3\arcsec, black line with error bars). The MIDI single-dish errors were taken as the error of the mean from five observations. Also plotted are blackbody emission curves at $T = 235, 285, 310$ K (green, red, blue). The region of atmospheric ozone absorption, between 9.3 and 10 $\mu$m (hatched), is uncertain and not taken into account for the later analysis.}
		\label{fig:phots}
	\end{centering}
\end{figure}

The resulting single-dish spectrum with an effective aperture of $\sim$ 300~mas is shown in Fig.~\ref{fig:phots} together with a Spitzer IRS spectrum for this source \citep[observed 8 Apr 2004; aperture $\sim 3.5$\arcsec,][]{weedman2005}. The higher Spitzer flux most likely indicates emission from the Narrow Line Region \citep[cf.][]{radomski2003}.

The color temperature of the spectrum is ($285^{+25}_{-50}$) K. Additionally, we detect the [Ne II] 12.81 $\mu$m line commonly seen in star forming regions. The [S IV] (10.51 $\mu$m) line, clearly seen in the Spitzer spectrum, is not significant in the MIDI spectrum.  The errors in the MIDI spectrum arise from incomplete thermal background subtraction and hence rise steeply with increasing wavelength.

\subsection{Correlated spectra}
\begin{figure}
	\begin{centering}
		\includegraphics[width=9.5cm,trim=3cm 2cm 0cm 3cm]{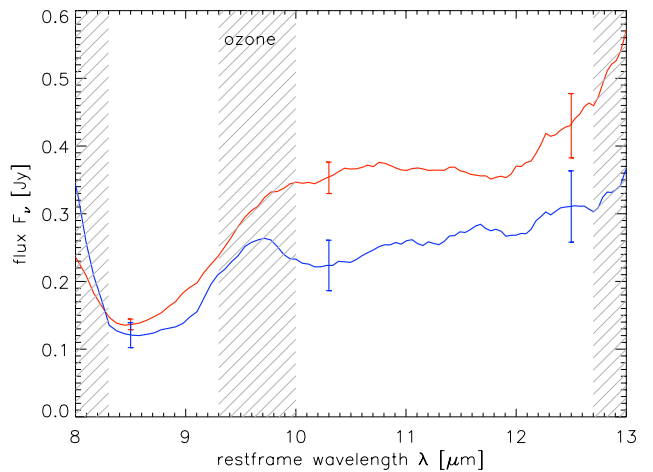}
		\caption{Smoothed ($\Delta \lambda \sim 0.3 \mu$m) correlated MIDI spectra at two different East-West projected baselines: 61m (red, average of two fringe track observations) and 89m (blue, average of three fringe track observations). The errors are the errors of the mean of the individual observations. The region of atmospheric ozone absorption, between 9.3 and 10 $\mu$m (hatched), is uncertain and not taken into account for the later analysis, as are the regions $\lambda < 8.3 \mu$m and $\lambda > 12.7 \mu$m (hatched), which have very low S/N.}
		\label{fig:corr}
	\end{centering}
\end{figure}

The two correlated flux spectra are shown in Fig.~\ref{fig:corr}.

The correlated flux observed with $BL = 61{\mathrm m}$ rises from $(0.13 \pm 0.01)$ Jy at 8.5 $\mu$m to $(0.43 \pm 0.05)$ Jy at 12.5 $\mu$m. The second spectrum ($BL = 89{\mathrm m}$), rises from $(0.12 \pm 0.02)$ Jy at 8.5 $\mu$m to $(0.30 \pm 0.05)$ Jy at 12.5 $\mu$m. Correlated flux uncertainties arise primarily from background photon noise and increase with wavelength but are smaller than the single dish errors.

The correlated flux on the shorter baseline (the one that has a higher flux) shows a broad ``bump'' between 9 $\mu$m and 12 $\mu$m that we interpret as a silicate emission feature (see section \ref{subsec:si}). No [Ne II] emission is seen, indicating that this arises on a scale that is fully resolved by the interferometer (i.e. $\gtrsim 20$ mas $\sim$ 1.3 pc). The fact that the correlated flux is lower on the longer baseline is a clear sign that the source is resolved by the interferometer.

Since the resolution of the interferometer $\theta_{min} \sim \lambda/3 BL$ changes with wavelength, the correlated flux reflects both the source spectrum and the source structure. It is not possible to draw any conclusions from the spectral slope of a correlated flux spectrum without assuming a source geometry.

%
%

\subsection{A possible silicate emission feature}
\label{subsec:si}

Although the silicate absorption feature has often been detected in type 2 nuclei, the emission feature, predicted for type 1 nuclei by torus models \citep[e.g.][]{pier1992,schartmann2005}, has not been detected except in a handful of objects, most of them quasars \citep{hao2005,weedman2005,buchanan2006}.

\begin{figure}[t]
	\begin{centering}
		\includegraphics[width=9.5cm,trim=3cm 2cm 0cm 2cm]{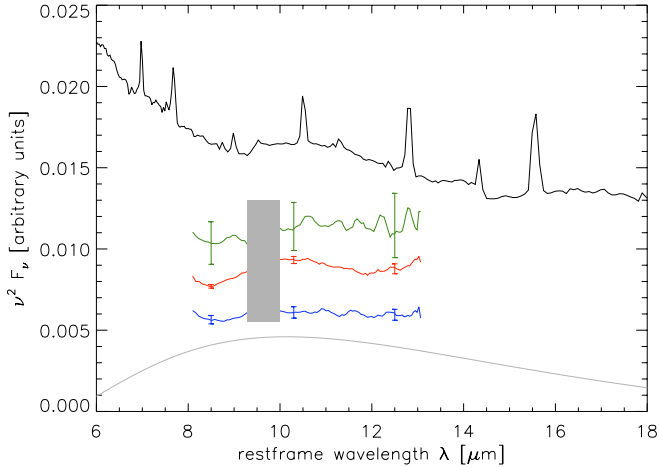}
		\caption{The silicate emission feature as seen in $\nu^2 F_{\nu}$. In the Spitzer spectrum \citep[black,][]{weedman2005}, a broad weak emission feature at around 11 $\mu$m is clearly seen. The three MIDI spectra (green: single-dish, red: 61m baseline, blue: 89m baseline) are plotted with offsets. The grey line shows a 285 K blackbody. The region between 9.3$\mu$m and 10 $\mu$m is hard to calibrate in the MIDI spectra due to the atmospheric ozone feature and has been greyed to not mislead the eye.}
		\label{fig:si}
	\end{centering}
\end{figure}

As noted by \citet{weedman2005} and \cite{buchanan2006}, Spitzer spectra (with an aperture of $\sim 3.5$\arcsec) of NGC 4151 show weak excess emission at 10 $\mu$m and 18 $\mu$m that is most easily seen when plotted as $\nu^2 F_{\nu}$ with a sufficiently large wavelength range. In Fig.~\ref{fig:si} we plotted our spectra (that are limited to the atmospheric N band) in such a way together with a Spitzer spectrum.

The hypothesis of silicate emission is not inconsistent with our single dish spectrum (aperture $\sim$ 0.3\arcsec) but since this spectrum suffers from incomplete background subtraction, especially at longer wavelengths, it is probably hidden in the resulting uncertainties. The emission feature seems to be most prominent in our observations on the 61m baseline observation and is clearly not detected on the 89m baseline.

This suggests that silicate emission is distributed over $\sim$ 1 pc as derived from the 10.5 $\mu$m Gaussian model (see \ref{sec:gauss}).

\subsection{Simple Gaussian model}
\label{sec:gauss}

With the limited baselines available from Paranal, it is not possible to reconstruct an {\it image} of NGC~4151 from our data.  Instead we consider simple model distributions of the emission on the sky and compare the predicted interferometric and single dish spectra with our measurements in order to fix parameters in such a model. We chose a model containing an unresolved point source (flux $F_p$) and an extended Gaussian distribution (flux normalization $F_g$, FWHM $\theta$). Although we might expect the mid-IR brightness distribution in Sy 1 galaxies to have a hole in the middle, neither observed nor modelled fluxes change as long as the hole, whose radius is determined by the sublimation radius of the dust, is unresolved. This is certainly the case in NGC 4151 (see section \ref{sec:dis}). An upper limit to the size of our point source is given by the effective resolution of the interferometer; this corresponds to a diameter of $\sim \lambda/3 BL \sim 7$ mas, i.e. a radius of $\sim$ 0.2 pc, at 10.3 $\mu$m.

Because of the East-West baseline orientation, the North-South distribution of emission is undetermined. Equivalently, we assume the source to be circularly symmetric on the sky.

\begin{figure}[t]
	\begin{centering}
		\includegraphics[width=9.5cm,trim=3cm 2cm 0cm 2cm]{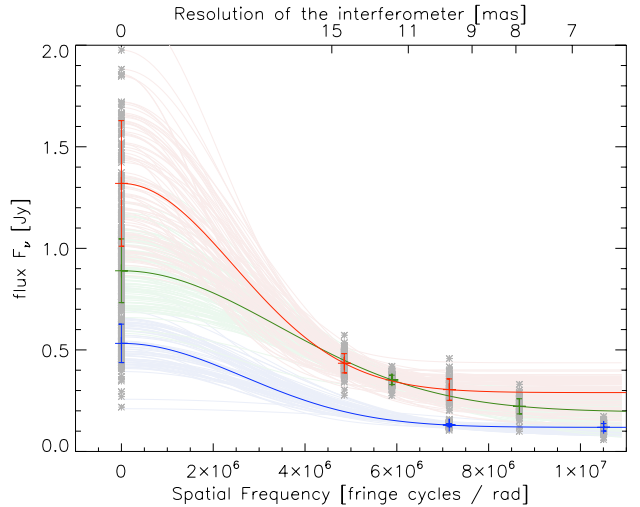}
		\caption{Gaussian model for NGC~4151. On the upper $x$ axis the resolution (i.e. sensitivity to model parameters) of the interferometer $\lambda / 3 BL$ is given. The three curves correspond to uniquely determined models at 8.5 $\mu$m (blue), 10.3 $\mu$m (green) and 12.5 $\mu$m (red). Data points and errors are taken from the single-dish (i.e. ``0-baseline'', Fig.~\ref{fig:phots}) and correlated fluxes (the remaining two data points per wavelength, Fig.~\ref{fig:corr}). The many thin lines represent the Monte-Carlo-like error determination for the Gauss model. For details see text.}
		\label{fig:model}
	\end{centering}
\end{figure}

The model correlated flux density at wavelength $\lambda$ is then given by $F_{\nu}(\lambda) = F_p(\lambda) + F_g(\lambda) \cdot \exp(- (BL/\lambda \cdot \pi/2 \cdot \theta)^2/\ln(2))$. With the given data points this model is uniquely determined. We calculated the parameters of such a model separately at 8.5, 10.3 and 12.5 $\mu$m where we are safely away from the regions of very low signal to noise (at the edges of the N band) and the ozone feature. At 10.3 $\mu$m the parameter values may be affected by the silicate feature. The modelled visibilities are shown in Fig.~\ref{fig:model} and the resulting parameters are given in Table \ref{tab:results}.

The errors of these parameters were estimated from a Monte-Carlo simulation in which we randomly placed 10000 measurements in a Gaussian distribution around the measured value with the $\sigma$ as determined from our data reduction. The parameter errors are then given by the standard deviations of the resulting model solutions to the simulated data.

%

\begin{table}
\caption{Parameters for the Gaussian models (see text).}
\begin{center}
\begin{tabular}{|c|c|c|c|}
\hline
$\lambda$ [$\mu$m]           & $F_g$ [Jy]         & $\theta$ [mas] (diameter [pc])         & $F_p$ [Jy] \\
\hline
$8.5 \pm 0.2$                &  $0.41 \pm 0.10$   &  $29^{+3} \mbox{}_{-6}$ ($2.0^{+0.2} \mbox{}_{-0.4}$) & $0.119 \pm 0.016$\\
$10.3 \pm 0.25$              &  $0.70 \pm 0.16$   &  $23 \pm 4$ ($1.5 \pm 0.3$) & $0.194 \pm 0.061$\\
$12.5 \pm 0.3$               &  $1.03 \pm 0.30$   &  $32 \pm 6$ ($2.1 \pm 0.4$) & $0.290 \pm 0.070$\\
\hline
\end{tabular}
\end{center}
\label{tab:results}
\end{table}

\section{Discussion}
\label{sec:dis}

\subsection{The extended source and the Sy 1 / Sy 2 paradigm}

In the strictest version of unified models, we expect for both a Sy 1 and a Sy 2 galaxy an extended dust structure with the same size, morphology and temperature distribution (at a given UV luminosity $L_{\rm UV}$). In less strict versions this is only true statistically \citep[e.g.][]{elitzur2006b}. Additionally, Sy 1 galaxies should have an unobscured point source (the unresolved accretion disk and inner rim of the torus) -- but the relative strength of these two components in the mid-IR is model-dependent.

To test this, we can compare our observations with other MIDI observations: In the Circinus galaxy ($L_{\rm UV} \sim 4 \times 10^{36}$ W, $L_{\rm torus} \sim 5 \times 10^{35}$ W), \citet{tristram2007b} found a warm ($T \sim 330$ K) disk with major axis FWHM $\sim 0.4$ pc and a larger, similarly warm, component of $\sim 2$ pc FWHM.

In NGC 1068 ($L_{\rm UV} \sim 3 \times 10^{37}$ W, $L_{\rm torus} \sim 10^{37}$ W), \citet{raban2009} found a hot ($T \sim 800$ K) disk of 1.35 x 0.45 pc and a warm component 3 $\times$ 4 pc in FWHM. They identified the disks with the densest parts of the torus of the unified model.

For NGC 4151 ($L_{\rm UV} \sim 1.5 \times 10^{36}$ W -- variable (from NED), $L_{\rm torus} \sim 4 \pi D^2 \nu F_g \sim 6 \times 10^{35}$ W) we determined a torus size (FWHM) of $\sim (2.0 \pm 0.4)$ pc and a dust temperature of $(285^{+25}_{-50})$ K.

When scaled to the accretion disk luminosity $L_{\rm UV}$, these values agree well with the torus sizes ($r \sim L_{\rm UV} \mbox{}^{1/2}$) and torus luminosities ($L_{\rm torus} \sim L_{\rm UV}$) of the two Seyfert 2 galaxies and the temperatures are also very similar. Note, however, that the torus luminosity used here is calculated from the $12 \mu$m flux density, not taking into account emission at longer wavelengths.

\subsection{Greybody models and the nature of the extended source}

In addition to the single-wavelength Gaussian models discussed above, we can also connect the various wavelengths together by constructing {\it greybody} models where we assume a smooth powerlaw dependence of temperature with radius: $T(r) \propto r^{-\alpha}$, and an essentially constant emissivity with radius and wavelength. With such models we get acceptable fits (reduced $\chi^2 \sim 1$) to both the total and the correlated flux spectra with $T({\rm 1 pc})\sim 250 $ K and $0.35 \lesssim \alpha \lesssim 0.6$. These values of $\alpha$ are consistent with dust receiving direct radiation from a central source \citep[e.g.][]{barvainis1987}, i.e. an optically thin medium with optically thick clumps in it.

These models can be extrapolated to shorter wavelengths to check their consistency with the K band measurements. All plausible extrapolations of the MIDI data yield K band fluxes of $< 10$ mJy, much less than observed by \citet{swain2003} and \citet{riffel2009} (see below). On the basis of these greybody models, we therefore conclude that the K band emission arises from structures which can probably not be extrapolated from the larger structure seen in the N band.

From the greybody models one can further infer an emissivity $\epsilon \sim 10^{-1}$ -- similar to what has been seen in NGC 1068 and Circinus.

To conclude: The resolved nuclear mid-IR structure in NGC 4151 has a size, temperature and emissivity that is comparable to those in type 2 objects where the existence of clumpy tori is established. Apart from the similarities with type 2 tori and the temperature profile of the greybody models, there is further evidence for clumpiness in the NGC 4151 torus from radio observations: \citet{mundell2003} measured HI absorption against the radio jet ($PA \sim 77 ^{\circ}$) and found a structure of $\sim 3$ pc in size. From the velocities they further suggest that the gas is distributed in clumps. Warm dust possibly traces the HI gas in the mid-IR.

It therefore seems reasonable to identify the warm dust structure resolved now in NGC 4151 with the clumpy tori seen in Sy 2 galaxies. Due to the limited observing geometry and the limited amount of observations we cannot reconstruct its apparent shape nor can we constrain a model with more than the two components discussed above.

\subsection{The point source and its relation to K band measurements}

K band interferometry measurements by \citet{swain2003} and, with higher significance, more recent Keck observations (Pott et al. 2009, in prep.) revealed a marginally resolved source, compatible with hot dust at the sublimation radius of $\sim 0.05$ pc. Reverberation measurements by \cite{minezaki2004} and \citet{koshida2009} find lag times $\Delta t$ between the UV/optical continuum and the K band corresponding to a radius of $\sim 0.04$ pc (variable) which they interpret as the sublimation radius of dust.

\citet{riffel2009} measured a flux of $\sim 65$ mJy at 2.2 $\mu$m and from spectral fitting they found that, at 2.2 $\mu$m, this flux is entirely dominated by a blackbody with a temperature of $(1285 \pm 50)$ K -- again consistent with hot dust at the sublimation radius and the K band interferometry, taking into account that $L_{\rm UV}$ is variable by at least a factor of 10 \citep{ulrich2000}. To account for that variability when comparing our observations with these K band observations of different epochs, we looked at the X-Ray flux as a proxy for the UV--optical radiation\footnote{Data from the All-Sky-Monitor (ASM) on the Rossi X-Ray Timing Explorer (RXTE), available online at http://xte.mit.edu/asmlc/srcs/ngc4151.html. The lag due to the light travel time $\sim c \Delta t \sim 48$ days was taken into account.} and find that, at the date of our observations, the source was probably in a higher state than at the time of \citet{riffel2009}'s observations and emitted $\sim 0.2$ Jy in the K band.

Since the flux density $F_{\nu}$ of a $\sim 1285$ K blackbody is roughly the same at $2.2 \mu$m as at 8.5 $\mu$m, we can compare the flux in the K band ($\sim 0.2$ Jy at the time of our measurement) with our point source flux at 8.5 $\mu$m ($\sim (0.119 \pm 0.016)$ Jy). From this it seems likely that hot dust is contributing to our point source at 8.5 $\mu$m. The spectrum of the point source rises by more than a factor of two from 8.5 $\mu$m to 12.5 $\mu$m, however, and thus requires emission from an additional small, ``red'' component. This could be core synchrotron emission, although such sources usually have flat spectra, or emission from a small, cool, optically thick central dust structure, possibly shadowed from direct accretion disk radiation.

\section{Conclusions}
Using mid-IR interferometry, we have resolved a warm dusty structure in NGC~4151. Its FWHM size ($2.0 \pm 0.4$ pc -- from comparing the data with a Gaussian model), temperature ($285^{+25}_{-50}$ K) and emissivity ($\sim 0.1$) are in good agreement with the clumpy tori seen in type 2 AGNs and are thus consistent with the unified model of Active Galaxies. Excess emission around 10.5 $\mu$m on the intermediate baseline indicates that silicate emission might be distributed out to $\sim$ 1 pc in AGNs.

Using simple models we compare our mid-IR fluxes with observations in the K band and find that the structure we resolve is probably not the smooth continuation of the nuclear source detected in the K band

Due to the limited number of measurements, no two-dimensional information could be gathered and questions about the unified model (such as: is the dust structure in Sy 1 galaxies thick and torus-like or rather disk-shaped?) remain unanswered. Since nuclear dust distributions are different even within the same class of AGNs, the ultimate question whether or not the unified model is valid will not be answered before a statistical study of numerous tori is performed.
\section*{Acknowledgements}
The authors wish to thank the anonymous referee for many comments that helped to improve the paper. LB wants to thank Roy van Boekel and Marc Schartmann for helpful discussions. DR wants to thank the Netherlands Organisation of Scientific Research (NWO) through grant 614.000.414.
\footnotesize
\bibliographystyle{apj}
\end{document}